\documentclass[trackchanges]{aastex701}

\newcommand{\kms}{\ensuremath{{\rm km\ s^{-1}}}}
\newcommand{\cmc}{\ensuremath{{\rm cm^{-3}}}}
\newcommand{\visc}{\ensuremath{{\rm g\ cm^{-1}\ s^{-1}}}}
\newcommand{\msp}{\ensuremath{\mu_{\rm sp}}}

\begin{document}

\title{A Kinematic Measurement of the Effective Viscosity of the Intracluster
Medium: The Cold Front of A3667 Seen by XRISM}

\author[orcid=0000-0003-0058-9719,sname='Fujita']{Yutaka Fujita}
\affiliation{Department of Physics, Graduate School of Science, Tokyo Metropolitan University, 1-1 Minami-Osawa, Hachioji-shi, Tokyo 192-0397, Japan}
\email[show]{y-fujita@tmu.ac.jp}

\author[orcid=0009-0009-9196-4174,sname='Omiya']{Yuki Omiya}
\affiliation{Institute of Space and Astronautical Science, Japan Aerospace Exploration Agency, 3-1-1 Yoshinodai, Chuo-ku, Sagamihara, Kanagawa 252-5210, Japan}
\email{ohmiya.yuuki@jaxa.jp}

\begin{abstract}
Constraints on the effective viscosity of the intracluster medium (ICM) come
almost exclusively from imaging: the power spectrum of X-ray surface brightness
fluctuations, and the morphology of Kelvin--Helmholtz rolls at cold fronts.
Both indicate a viscosity below the isotropic Spitzer value. XRISM/Resolve
opens a complementary, purely kinematic route, in which the viscosity is
inferred from the thickness of the viscous shear layer that develops at a
sliding contact discontinuity. We apply it to the prototypical cold front of
A3667. The interface is the classical Rayleigh problem, in which momentum
diffuses to a depth $d$ in a time $T$, giving a kinematic viscosity
$\nu \approx d^{2}/(4T)$. The XRISM velocity map shows a line-of-sight velocity
jump of $535$~\kms\ across the front, completed within the $100$~kpc-wide
region immediately inside it, the only region with an anomalously broad line.
Hence $d \lesssim 100$~kpc. Adopting interaction times set by the sizes of the
interacting regions divided by this velocity, we obtain dynamic viscosities of
$5.2 \times 10^{3}$ and $3.3 \times 10^{3}$~\visc\ for the cool and the hot
side of the front, or $6.4$ and $0.72$ times the Spitzer value. Because $d$ is
limited by the angular resolution of XRISM, and the viscosity scales as its square, these
are upper limits. The kinematic constraint is consistent with, and independent
of, the imaging results.
\end{abstract}

\keywords{\uat{Galaxy clusters}{584} --- \uat{Intracluster medium}{858} --- \uat{Plasma astrophysics}{1261} --- \uat{High resolution spectroscopy}{2096}}

\section{Introduction}
\label{sec:intro}

The intracluster medium (ICM) is a hot, dilute, magnetized plasma whose
microphysics is still largely unconstrained. Among the transport coefficients,
the shear viscosity is of particular interest, because it controls the
dissipation scale of the turbulent cascade, the survival of cold fronts and
stripped tails, and the amount of non-thermal pressure support that biases
cluster mass estimates. For a collisional, unmagnetized plasma the dynamic
viscosity is given by the classical Spitzer value \citep{1962pfig.book.....S},
which for ICM conditions is a steep function of temperature. In a magnetized,
weakly collisional plasma, however, the momentum transport is anisotropic with
respect to the local field \citep{1965RvPP....1..205B}, and it may be further
suppressed by the scattering of particles off microscale fluctuations driven by
pressure-anisotropy instabilities \citep{2014PhRvL.112t5003K}. The effective
viscosity relevant on macroscopic scales is therefore an empirical quantity
that must be measured.

Almost all existing measurements are based on images. The power spectrum of
X-ray surface brightness fluctuations in the Coma cluster extends to scales at
or below the Coulomb mean free path without the steepening that a Spitzer-like
viscosity would produce, requiring an effective viscosity well below the
isotropic Spitzer value \citep{2019NatAs...3..832Z}. Independently, the
appearance --- or the absence --- of Kelvin--Helmholtz (KH) rolls at cold
fronts has been compared with viscous hydrodynamic simulations
\citep{2013MNRAS.436.1721R,2013ApJ...764...60R}, again pointing to a strongly
suppressed viscosity in several systems
\citep{2017MNRAS.467.3662I,2021MNRAS.504.2800I}; for A3667,
\citet{2017MNRAS.467.3662I} place the effective viscosity at no more than
$5$ per cent of the Spitzer value. All of these arguments, however, infer the
velocity field indirectly, from the morphology of the gas density distribution.

The situation has changed with XRISM \citep{2025PASJ...77S...1T}. Following
Hitomi \citep{2016Natur.535..117H}, the Resolve microcalorimeter now measures
line-of-sight (LoS) bulk velocities and velocity dispersions of the ICM
directly and with an accuracy of tens of \kms\
\citep[e.g.,][]{2025Natur.638..365X}. In particular, \citet{2026ApJ...996L..15O}
have mapped the velocity field across the prototypical cold front of A3667 and
found a LoS velocity jump of $535^{+167}_{-154}$~\kms\ across the interface,
accompanied by a strongly enhanced LoS velocity dispersion,
$424^{+125}_{-80}$~\kms, in the region immediately inside the
front. These are quantities that a viscous shear layer would produce. We do not
assume that the whole of the observed dispersion is due to shear; instead we
use the shear contribution it can at most account for, which is what yields an
upper limit on the thickness of the layer and hence on the viscosity.

In this paper, we demonstrate that these measurements enable a kinematic determination of the effective viscosity, which complements the aforementioned imaging methods. Our approach is straightforward. It involves the classic problem of a viscous fluid being dragged by a plate that accelerates suddenly. We apply this method to A3667. Section~\ref{sec:model} presents the model,
Section~\ref{sec:a3667} applies it to A3667, and Sections~\ref{sec:result}
and \ref{sec:discussion} give the results and discuss their limitations.
Throughout we adopt the cosmology used by \citet{2026ApJ...996L..15O},
$H_{0} = 70$~\kms~Mpc$^{-1}$, $\Omega_{m} = 0.3$ and
$\Omega_{\Lambda} = 0.7$; at the cluster redshift $z = 0.055$, $1' $
corresponds to $64$~kpc.

\section{Viscous Shear Layer at a Sliding Interface}
\label{sec:model}

Consider two bodies of gas in contact along a plane interface, one of which
starts to move tangentially with a constant speed $U$ relative to the other at
$t = 0$. Let $y$ be the coordinate perpendicular to the interface and $u$ the
velocity component parallel to it. If the flow is incompressible and there is
no pressure gradient along the interface, the Navier--Stokes equation reduces
to the diffusion equation
\begin{equation}
\frac{\partial u}{\partial t} = \nu \frac{\partial^{2} u}{\partial y^{2}} \, ,
\label{eq:ns}
\end{equation}
where $\nu = \mu / \rho$ is the kinematic viscosity, $\mu$ the dynamic
viscosity and $\rho$ the gas density. For the initial and boundary conditions
\begin{eqnarray}
u(y) &=& 0 \quad {\rm at} \quad t = 0 \, , \\
u &=& U \quad {\rm at} \quad y = 0 \quad {\rm for} \quad t > 0 \, ,
\end{eqnarray}
together with $u \rightarrow 0$ as $y \rightarrow \infty$, the solution of
equation~(\ref{eq:ns}) is the well-known self-similar profile
\begin{equation}
u = U \left[ 1 - {\rm erf}\, \eta \right] \, , \qquad
\eta = \frac{y}{2 \sqrt{\nu t}} \, ,
\label{eq:erf}
\end{equation}
where ${\rm erf}\, \eta = (2/\sqrt{\pi}) \int_{0}^{\eta} e^{-s^{2}} {\rm d}s$
is the error function. This is Stokes' first problem, or the Rayleigh problem
of an impulsively started plate.

Equation~(\ref{eq:erf}) shows that the velocity perturbation is confined to a
layer whose thickness is set by $\eta \sim 1$: at $\eta = 1$ the velocity has
dropped to $u \approx 0.16\,U$, and for $\eta > 1$ the gas is essentially undisturbed.
Identifying the observed depth of the dragged layer with $y = d$ and the
elapsed interaction time with $t = T$, the condition $\eta = 1$ gives
\begin{equation}
\nu \approx \frac{d^{2}}{4 T} \, ,
\label{eq:nu}
\end{equation}
which is the basic relation used below. The gas mass density follows from the
electron number density as $\rho = \mu_{e} m_{p} n_{e}$, with
$\mu_{e} \approx 1.17$ for a fully ionized plasma and
$m_{p}$ the proton mass, so that the dynamic viscosity is
\begin{equation}
\mu = \rho\, \nu \approx \frac{\mu_{e} m_{p} n_{e} d^{2}}{4 T} \, .
\label{eq:mu}
\end{equation}
The only two quantities that need to be supplied by observation are therefore
the thickness $d$ of the shear layer and the duration $T$ over which the two
bodies of gas have been in contact. Both are, in principle, directly accessible
to a high-resolution spectrometer: $d$ from the spatial extent of the region
displaying an anomalously broad line, and $T$ from the geometry of the flow
divided by the shear velocity.

One point of definition should be noted from the beginning. In
equation~(\ref{eq:erf}) $U$ is a boundary condition: the relative velocity of
the two bodies of gas {\it far from} the interface, where they have not yet
been dragged. A spectrometer, by contrast, measures velocities averaged over
finite regions, each of which contains part of the shear layer, so the
difference between two region-averaged velocities underestimates $U$. Since
$T \propto U^{-1}$ and $\mu \propto T^{-1}$, using the measured difference in
place of $U$ underestimates $\mu$.

This sets the division of labor for what follows.
Sections~\ref{sec:a3667} and \ref{sec:result} take the one-sided
equation~(\ref{eq:erf}) at face value, treating the measured velocities as if
they described a simple two-valued distribution across the front, and obtain
the viscosities from equation~(\ref{eq:mu}) directly. Section~\ref{sec:discussion}
then relaxes that idealization: it develops the two-sided form of
equation~(\ref{eq:erf}), in which both bodies of gas are dragged and the
interface itself moves, and uses it to quantify the systematic uncertainties
that the simple treatment leaves out --- the underestimate of $U$ just
described, and the composite line profiles that arise when the shear layer is
unresolved or when emission from one region leaks into another.

\section{Application to the Cold Front of A3667}
\label{sec:a3667}

\subsection{Geometry}
\label{sec:geometry}

A3667 hosts one of the sharpest and most extensively studied cold fronts
\citep{1999ApJ...521..526M,2001ApJ...551..160V,2007PhR...443....1M}.
\citet{2026ApJ...996L..15O} observed the front and the cluster core with
XRISM/Resolve and divided the field into seven sky regions (their regions
A--G). Region A lies outside the front, on the less dense side; region B is the
bright, cool region immediately inside it; regions C and D lie further inside;
and regions E--G cover the central ICM. Throughout this paper $\sigma_{z}$
denotes a LoS velocity dispersion. For the values measured by
\citet{2026ApJ...996L..15O} we write $v_{\rm X}$ for the LoS bulk velocity of
region X relative to the brightest cluster galaxy (BCG), and $\sigma_{\rm X}$
for its LoS velocity dispersion; the same subscript convention is used for
other quantities. Across the front,
\begin{equation}
v_{\rm A} = -317^{+102}_{-108}~\kms \, , \qquad
v_{\rm B} = +218^{+128}_{-115}~\kms \, ,
\label{eq:vavb}
\end{equation}
a region-averaged velocity difference of
$v_{\rm B} - v_{\rm A} = 535^{+167}_{-154}$~\kms. The velocity dispersion is
$\sigma_{\rm X} \approx 150$--$200$~\kms\ in essentially all regions except
region B, where it rises to $\sigma_{\rm B} = 424^{+125}_{-80}$~\kms. We
adopt $U = 535$~\kms\ as our fiducial shear velocity, bearing in mind that it
is a lower bound on the far-field relative velocity, for the reason given in
Section~\ref{sec:model}; the size of the systematic is quantified in
Section~\ref{sec:sysU}.

\begin{figure*}[ht!]
\plotone{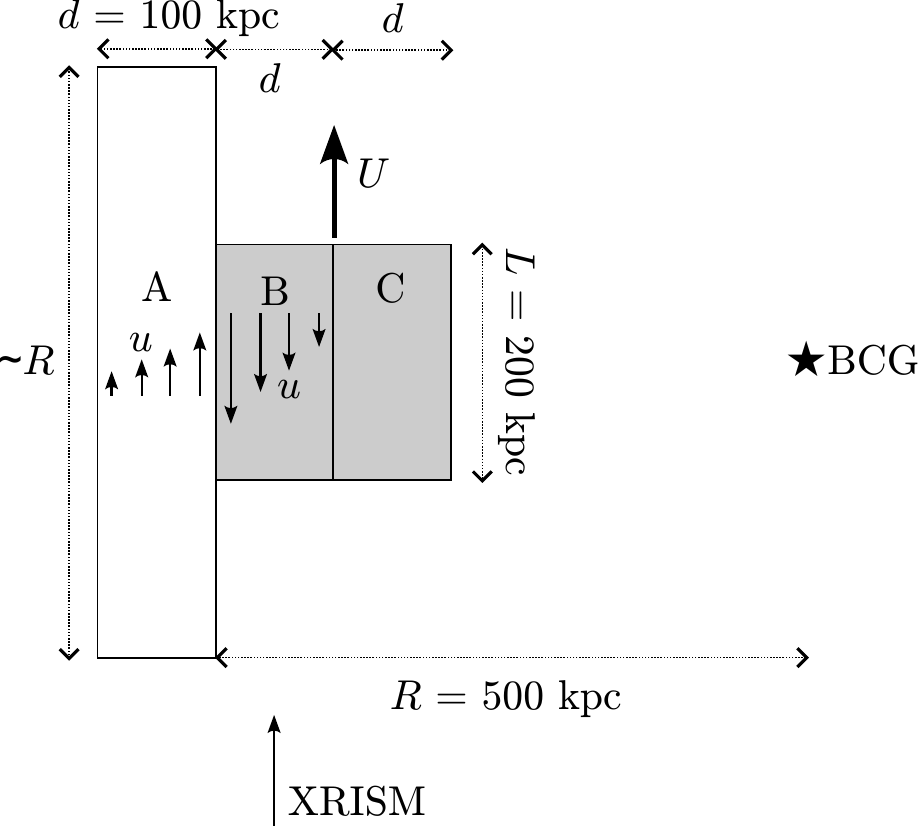}
\caption{Schematic of the shear flow at the A3667 cold front, drawn in the
plane that contains the line of sight; XRISM observes from the bottom. The cool
low-entropy gas forming regions B and C is approximated by a cube of side
$L = 200$~kpc that moves along the line of sight with a velocity $U$ relative
to the ambient hot gas of region A. The interface between A and the cube is
parallel to the line of sight, so the relative motion is tangential to it.
Viscous drag produces a boundary layer of thickness $d \approx 100$~kpc on each
side, occupying region B on the cool side and region A on the hot side; the
short arrows show the velocity profiles of equation~(\ref{eq:erf}) in the rest
frame of the gas on the other side of the interface. The line-of-sight depth of
region A is $\sim R \approx 500$~kpc, comparable to its projected distance from
the BCG.
\label{fig:geometry}}
\end{figure*}

\citet{2026ApJ...996L..15O} interpret these data as an offset merger in which
the low-entropy core gas is sloshing in the plane nearly perpendicular to the
sky, so that the shear across the front is directed almost entirely along the
LoS. Figure~\ref{fig:geometry} shows a schematic, minimal representation of
this configuration in the plane containing the LoS. The cool, low-entropy gas
that forms regions B and C is idealized as a cube\footnote{In \citet{2026ApJ...996L..15O} the low-entropy
gas extends over regions B--D; we omit region D from the model for simplicity.}
of side $L = 200$~kpc that
slides along the LoS with speed $U$ relative to the ambient hot gas of region
A. The interface between region A and the cube is parallel to the LoS, so that
$U$ is tangential to it and the configuration is exactly that of
Section~\ref{sec:model}, with $y$ measured in the plane of the sky,
perpendicular to the interface. Viscous drag then generates a boundary layer on
each side of the interface: on the cool side it occupies region B, and on the
hot side it occupies region A. Small arrows in Figure~\ref{fig:geometry}
indicate the corresponding velocity profiles, drawn in the rest frame of the
gas on the opposite side of the interface.

\subsection{Thickness of the Shear Layer}
\label{sec:d}

Because the velocity gradient of the boundary layer lies in the plane of the
sky, a spectrum extracted from a sky region that straddles the layer is a
superposition of lines emitted by gas moving at LoS velocities ranging from
$\sim 0$ to $\sim U$. Such a region therefore shows an anomalously broad line
even when the underlying turbulent motions are weak. This is what is observed:
the velocity dispersion is enhanced in region B alone,
$\sigma_{\rm B} = 424^{+125}_{-80}$~\kms, and returns to the ambient
$150$--$200$~\kms\ in the adjacent regions C and D, only $\sim 100$~kpc further
in. The velocity transition is thus complete within region B, whose width in
the direction perpendicular to the interface is $\approx 100$~kpc, and we adopt
\begin{equation}
d \approx 100~{\rm kpc} \, .
\end{equation}

We emphasize that this value should be regarded as an upper limit, for two
distinct reasons.

First, the half-power diameter of the XRISM mirror is $\sim 1\farcm 3$, or
$83$~kpc at $z = 0.055$, comparable to the width of region B itself. A layer
appreciably thinner than $100$~kpc would produce a spectrum that is
indistinguishable, at the present angular resolution, from the one observed.

Second, the measured dispersion is in fact larger than a laminar shear layer
can produce. Any distribution of LoS velocities confined between $0$ and $U$
has a dispersion of at most $U/2 = 268$~\kms, a value attained only for an
unresolved step in which the two velocities contribute equally. The fully
developed profile of equation~(\ref{eq:erf}) falls well short of this bound:
averaging it with uniform weight in $y$ over a layer that exactly fills region
B, i.e., over $0 \le \eta \le 1$, gives
$\langle u \rangle = 0.514\,U$ and $\sigma_{z} = 0.249\,U \approx 130$~\kms.
This is a volume-weighted value. Weighting instead by the emissivity
($\propto n_{e}^{2} \Lambda(T)$) gives greater weight to the denser gas deeper
inside the front, where $u \rightarrow U$, which raises $\langle u \rangle$ and
lowers $\sigma_{z}$ further; the bound $\sigma_{z} \le U/2$, on the other hand,
holds for any weighting. Reproducing the observed $\sigma_{\rm B}$ therefore
requires a contribution beyond the projection of the mean shear profile ---
turbulence generated within the shear layer, unresolved larger-scale
merger-driven flows, or a true tangential velocity exceeding the LoS-projected
$535$~\kms\ if the motion is inclined to the LoS. This is the same conclusion
reached by \citet{2026ApJ...996L..15O}, who attribute the broadening to a
mixture of turbulent and bulk kinematic contributions. In each of these cases
the share of the broadening carried by the laminar profile is smaller than
observed, so the layer need not fill region B at all. One further possibility,
residual spectral leakage from the adjacent region A, is deferred to
Section~\ref{sec:leak}, since it interacts with the systematics of $U$
discussed there.

Both arguments point the same way: $d \le 100$~kpc, and since
$\nu \propto d^{2}$ the derived viscosities are upper limits. We adopt the same
$d$ for the layer on the hot side, where the dispersion of region A,
$\sigma_{\rm A} = 200^{+97}_{-200}$~\kms, is too poorly constrained to provide an independent
estimate.

\subsection{Interaction Timescales}
\label{sec:time}

The interaction time $T$ is the duration over which a given fluid element has
been in contact with gas moving at a different velocity. It differs between the
two sides of the interface because the two bodies of gas have different extents
along the direction of relative motion.

For the cool gas of region B, the relevant length is the LoS depth of the
ambient region A that it must traverse. Taking the LoS depth of region A to be
comparable to its projected distance from the BCG,
$R \approx 500$~kpc (Figure~\ref{fig:geometry}), we obtain
\begin{equation}
T_{\rm B} \approx \frac{R}{U} \approx 0.91~{\rm Gyr} \, .
\end{equation}
For the hot gas of region A, in contrast, the moving body that drags it is the
cube itself, whose extent along the LoS is $L = 200$~kpc. A fluid element of
region A is therefore in contact with the moving gas only for
\begin{equation}
T_{\rm A} \approx \frac{L}{U} \approx 0.37~{\rm Gyr} \, .
\end{equation}
Both timescales are shorter than the age of the merger, so the boundary layers
are still growing and the steady-state assumption is not required.

\section{Results}
\label{sec:result}

Substituting the values above into equations~(\ref{eq:nu}) and (\ref{eq:mu}),
and adopting the electron densities of the inner and outer sides of the front
derived from the Chandra surface brightness profile,
$n_{e} = 3.2 \times 10^{-3}$~\cmc\ for region B and
$8.2 \times 10^{-4}$~\cmc\ for region A \citep{2001ApJ...551..160V}, we obtain
\begin{eqnarray}
\nu_{\rm B} &=& \left( 8.3^{+2.6}_{-2.4} \right) \times 10^{29}~{\rm cm^{2}\ s^{-1}} \, , \qquad
\mu_{\rm B} = \left( 5.2^{+1.6}_{-1.5} \right) \times 10^{3}~\visc \, , \\
\nu_{\rm A} &=& \left( 2.1^{+0.6}_{-0.6} \right) \times 10^{30}~{\rm cm^{2}\ s^{-1}} \, , \qquad
\mu_{\rm A} = \left( 3.3^{+1.0}_{-1.0} \right) \times 10^{3}~\visc \, .
\end{eqnarray}
The quoted uncertainties propagate the measurement error on the velocity jump,
$U = 535^{+167}_{-154}$~\kms, which is the only quantity in
equation~(\ref{eq:mu}) for which \citet{2026ApJ...996L..15O} provide a
statistical error, and which enters linearly through $T \propto U^{-1}$. They
do not include the uncertainties on the electron densities, which
\citet{2001ApJ...551..160V} quote without formal errors for region~A, nor those on the
assumed lengths $R$ and $L$; and they are of course not applicable to $d$,
which is a bound rather than a measurement. These are collected in
Table~\ref{tab:results}.

For comparison, the Spitzer viscosity of a fully ionized hydrogen plasma is
\citep[][their equation~5.46]{1986RvMP...58....1S}
\begin{equation}
\msp \approx 5500
\left( \frac{k T}{8.62~{\rm keV}} \right)^{5/2}
\left( \frac{\ln \Lambda}{40} \right)^{-1} ~\visc \, ,
\label{eq:spitzer}
\end{equation}
where $8.62$~keV corresponds to $10^{8}$~K. Adopting $\ln \Lambda = 40$ and the
temperatures measured by \citet{2026ApJ...996L..15O},
$kT = 3.99^{+0.26}_{-0.22}$~keV for region B and $kT = 8.00$~keV (held fixed in
their fit) for region A, this gives $\msp = 8.0^{+1.4}_{-1.1} \times 10^{2}$
and $4.6 \times 10^{3}$~\visc, respectively. Combining the two sets of errors
in quadrature, the measured values correspond to
\begin{equation}
\frac{\mu_{\rm B}}{\msp} = 6.4^{+2.2}_{-2.2} \, , \qquad
\frac{\mu_{\rm A}}{\msp} = 0.72^{+0.23}_{-0.21} \, ,
\end{equation}
both of which are to be read as upper limits because of the bound on $d$.
On the cool side the kinematic estimate exceeds the Spitzer value by a factor
of a few --- though not by an order of magnitude, and only by
$2.5\sigma$ --- while on the hot side the two agree within the errors. Since
both numbers are upper limits, the data are equally compatible with
$\mu \ll \msp$.

\begin{deluxetable*}{lcccccccc}
\tablecaption{Kinematic viscosity estimates for the two sides of the A3667 cold
front\label{tab:results}}
\tablehead{
\colhead{Region} & \colhead{$n_{e}$} & \colhead{$kT$} & \colhead{$d$} &
\colhead{$T$} & \colhead{$\nu$} & \colhead{$\mu$} & \colhead{$\msp$} &
\colhead{$\mu/\msp$} \\
\colhead{} & \colhead{($10^{-3}$~\cmc)} & \colhead{(keV)} & \colhead{(kpc)} &
\colhead{(Gyr)} & \colhead{($10^{29}$~cm$^{2}$~s$^{-1}$)} &
\colhead{(\visc)} & \colhead{(\visc)} & \colhead{}
}
\startdata
B (inside the front)  & $3.2$  & $3.99^{+0.26}_{-0.22}$ & $100$ & $0.91^{+0.37}_{-0.22}$ & $8.3^{+2.6}_{-2.4}$  & $\left( 5.2^{+1.6}_{-1.5} \right) \times 10^{3}$ & $\left( 8.0^{+1.4}_{-1.1} \right) \times 10^{2}$ & $6.4^{+2.2}_{-2.2}$ \\
A (outside the front) & $0.82$ & $8.00$ (fixed) & $100$ & $0.37^{+0.15}_{-0.09}$ & $20.6^{+6.4}_{-5.9}$ & $\left( 3.3^{+1.0}_{-1.0} \right) \times 10^{3}$ & $4.6 \times 10^{3}$ & $0.72^{+0.23}_{-0.21}$ \\
\enddata
\tablecomments{Electron densities are from \citet{2001ApJ...551..160V} and
temperatures from \citet{2026ApJ...996L..15O}. The interaction times are
$T_{\rm B} = R/U$ and $T_{\rm A} = L/U$ with $R = 500$~kpc, $L = 200$~kpc and
the fiducial $U = 535^{+167}_{-154}$~\kms. Quoted errors propagate the
statistical uncertainties on $U$ and, for region B, on $kT$; the electron
densities, $R$, $L$ and $d$ carry no formal errors, and the temperature of
region A was held fixed in the spectral fit. Because $d$ is an upper limit set
by the angular resolution of XRISM and $\nu \propto d^{2}$, the tabulated
$\nu$, $\mu$ and $\mu/\msp$ are upper limits in that respect. They are, on the
other hand, underestimates with respect to $U$, which the region-averaged
velocity difference bounds from below (Section~\ref{sec:sysU}); this effect is
linear and at most a factor of two, and is not included in the quoted errors.}
\end{deluxetable*}

\section{Discussion}
\label{sec:discussion}

\subsection{The Shear Velocity as a Lower Bound}
\label{sec:sysU}

The velocity that enters equation~(\ref{eq:mu}) through $T$ is the far-field
relative velocity of the two bodies of gas, whereas what is measured is the
difference between two region-averaged velocities, each of which is pulled
toward the other by the boundary layer it contains. Quantifying the gap
requires the two-sided version of the problem of Section~\ref{sec:model}, in
which both bodies of gas are free to be dragged.

Let the interface lie at $y = 0$, with region A occupying $y < 0$ and at rest
far from the interface, and region B occupying $y > 0$ and moving at $U$ far
from it. The interface itself then moves at some intermediate velocity $u_{i}$,
and each side connects its own far field to the interface through a Stokes
solution,
\begin{equation}
u = u_{i}\, {\rm erfc} \left( \frac{|y|}{2 \sqrt{\nu_{\rm A} t}} \right)
\;\; (y < 0) \, , \qquad
u = u_{i} + (U - u_{i})\, {\rm erf} \left( \frac{y}{2 \sqrt{\nu_{\rm B} t}} \right)
\;\; (y > 0) \, .
\label{eq:twoside}
\end{equation}
The value of $u_{i}$ follows from the continuity of the viscous stress. Using
$d\,{\rm erf}(z)/dz = (2/\sqrt{\pi}) e^{-z^{2}}$, the velocity
gradients at $y \rightarrow 0$ are $u_{i}/\sqrt{\pi \nu_{\rm A} t}$ and
$(U - u_{i})/\sqrt{\pi \nu_{\rm B} t}$ on the two sides. Equating the stresses
$\mu\, \partial u / \partial y$, the factors of $\pi$ and $t$ cancel and
$\mu/\sqrt{\nu} = \sqrt{\mu \rho}$, so that
$\sqrt{\mu_{\rm A} \rho_{\rm A}}\, u_{i}
= \sqrt{\mu_{\rm B} \rho_{\rm B}} (U - u_{i})$, or
\begin{equation}
u_{i} = U \frac{\sqrt{\mu_{\rm B} \rho_{\rm B}}}
{\sqrt{\mu_{\rm A} \rho_{\rm A}} + \sqrt{\mu_{\rm B} \rho_{\rm B}}}
\approx 0.71\, U
\label{eq:ui}
\end{equation}
with the values of Table~\ref{tab:results}. That $u_{i}$ is independent of $t$
is required for self-similarity. The quantity $\sqrt{\mu \rho}$ acts as a
viscous impedance, in the same way that the thermal effusivity
$\sqrt{\kappa \rho c}$ --- with $\kappa$ the thermal conductivity and $c$ the
specific heat capacity --- governs two solids brought into thermal contact: the
side with the larger value yields less, and the interface moves with it. Note
also that $u_{i}$ is set mainly by the
density contrast rather than by the viscosities themselves: putting
$\mu_{\rm A} = \mu_{\rm B}$ gives
$u_{i} = U \sqrt{\rho_{\rm B}/\rho_{\rm A}} /
(1 + \sqrt{\rho_{\rm B}/\rho_{\rm A}}) \approx 0.66\,U$, so the estimate is not
materially circular even though the $\mu$ that enter it are the ones derived
here.

Averaging equation~(\ref{eq:twoside}) with uniform weight in $y$ over a region
that reaches $\eta = a$ into region A and one that reaches $\eta = b$ into
region B gives
\begin{eqnarray}
\langle u \rangle_{\rm A} &=& u_{i} \left[ 1 - E(a) \right] \, , \\
\langle u \rangle_{\rm B} &=& u_{i} + (U - u_{i}) E(b) \, ,
\label{eq:uAuB}
\end{eqnarray}
where
\begin{equation}
E(x) \equiv \frac{1}{x} \int_{0}^{x} {\rm erf}\, \eta \; {\rm d}\eta \, ,
\label{eq:Edef}
\end{equation}
with $E(1) = 0.486$ and $E \rightarrow 1$ for $x \gg 1$. The observable is the
difference,
\begin{equation}
\langle u \rangle_{\rm B} - \langle u \rangle_{\rm A}
= u_{i} E(a) + (U - u_{i}) E(b) \, .
\label{eq:average}
\end{equation}
Two limits are worth noting. If both regions are much wider than their layers
($a, b \gg 1$), the measured difference recovers $U$, as it should. If instead
each region is exactly as wide as its own layer ($a = b = 1$) --- the case
drawn in Figure~\ref{fig:geometry} --- the interface velocity cancels
identically and
$\langle u \rangle_{\rm B} - \langle u \rangle_{\rm A} = 0.486\, U$, so the
measurement underestimates $U$ by a factor of two. Setting the left-hand side
to the observed $\langle u \rangle_{\rm B} - \langle u \rangle_{\rm A} = 535$
\kms\ then gives $U = 1101$~\kms: both interaction times are halved and the
viscosities double, to $\mu_{\rm B} = 1.1 \times 10^{4}$ and
$\mu_{\rm A} = 6.8 \times 10^{3}$~\visc, or $13$ and $1.5$ times Spitzer. Since
$\mu \propto U$ while $\mu \propto d^{2}$, this linear systematic is
subdominant to the quadratic one attached to $d$, but it acts in the opposite
direction and should be kept in view.

The velocity map itself argues that the correction is closer to unity than to
two. If the layer filled region B, the mean velocity of region B would be
dragged away from that of the undisturbed cool gas further inside --- regions C
and D, which lie beyond the layer in the same body of gas --- and toward the
velocity of region A, which is to say toward the blue. The predicted shift is
$\langle u \rangle_{\rm B} - v_{\rm C} = -(1 - E(1))(U - u_{i})$, the
difference between equation~(\ref{eq:uAuB}) evaluated at $b = 1$ and at
$b \gg 1$, or $-164$~\kms\ for $U = 1101$~\kms. No such offset is seen: region
B is measured at
$v_{\rm B} = +218^{+128}_{-115}$~\kms\ against
$v_{\rm C} = +194^{+57}_{-71}$~\kms\ and $v_{\rm D} = +165^{+57}_{-37}$~\kms,
i.e., region B is if anything slightly
the more redshifted of the three. The observed difference,
$v_{\rm B} - v_{\rm C} = +24$~\kms, departs from the prediction by
$\sim 190$~\kms. Combining in quadrature the errors that would move the two
velocities toward the predicted shift --- the lower error on region B and the
upper error on region C, $\sqrt{115^{2} + 57^{2}} = 128$~\kms\ --- this is a
$1.5\sigma$ discrepancy, so the test is suggestive rather than decisive; a
sharper statement would require the joint confidence region for the two
velocities, which are fitted simultaneously and are therefore not independent.
Taken at face value, however, it indicates that the bulk of the gas in region B
has not been dragged, which requires the layer to be thinner than the region
--- and therefore reinforces, through the quadratic dependence, the reading of
$d \approx 100$~kpc as an upper limit.

\subsection{Spectral Leakage from the Adjacent Region}
\label{sec:leak}

Region B is separated from region A by less than the half-power diameter of the
XRISM mirror, so a natural worry is that its anomalously broad line is an
artifact of spectral leakage --- PSF scattering of region A photons into the
detector pixels assigned to region B. Such leakage is certainly present. In the
region b panel of Figure~2 of \citet{2026ApJ...996L..15O} the region A model
contributes at the level of $\sim 8\%$ of the region B model at the peak of the
Fe~K$\alpha$ complex, and its own peak lies blueward of it, as expected for gas
that is blueshifted relative to region B.

Leakage of this kind does not, however, inflate the measured $\sigma_{\rm B}$,
because the SSM analysis does not treat the region b spectrum as a single line.
Each sky region enters with its own \texttt{bapec} component, carrying its own redshift
and its own Gaussian width, and the cross-region ARFs fix how much of each
component lands on which detector pixels. The value
$\sigma_{\rm B} = 424^{+125}_{-80}$~\kms\ is therefore a parameter of the region B
component after the region A contribution has been accounted for, not a width
fitted to the blend. Blending is excluded here by construction rather than by
argument.

What remains is the possibility that the leakage fraction is misestimated, so
that some region A flux is absorbed into the region B component. The
arithmetic of Section~\ref{sec:d} bounds this. Leaked emission enters at
$u \approx 0$, one endpoint of the allowed interval, so the composite remains a
mixture confined to $[0, U]$ and a component of fractional weight $f$
contributes $\sigma_{z} = U \sqrt{f(1-f)}$. Even if the {\it entire} region A
contribution seen in Figure~2 of \citet{2026ApJ...996L..15O} were unmodeled,
$f \approx 0.08$ gives $144$~\kms\ for the fiducial $U = 535$~\kms\ and
$296$~\kms\ for the maximally corrected $U = 1101$~\kms\ of
Section~\ref{sec:sysU} --- in both cases short of $424$~\kms, and the true
unmodeled fraction is a small residual of the $8\%$, not the whole of it.

The measured width thus continues to require a contribution beyond both the
laminar shear profile and any plausible leakage, most naturally genuine
turbulence within the layer. We note in passing that the velocity range spanned
by the gas of region B on its own is not $U$ but $U - u_{i}$, the smaller share
of the shear by equation~(\ref{eq:ui}); for $U = 1101$~\kms\ this is
$317$~\kms, so the laminar contribution to an unblended region B line cannot
exceed $159$~\kms. Whichever source dominates, the conclusion for our purpose
is unchanged: a contribution from anything other than the laminar profile
reduces the share of the broadening that profile must carry, so that the layer
need not fill region B, and $d$ --- and with it $\mu$ --- is lowered still
further.

\subsection{Other Systematic Uncertainties}

The dominant uncertainty is the layer thickness $d$, which enters quadratically.
As discussed in Section~\ref{sec:d}, the value $d \approx 100$~kpc is comparable
to the point spread function of XRISM and to the width of the smallest sky
region for which spectra could be extracted, so it is a resolution-limited
upper bound rather than a measurement. A second, related uncertainty is the
inclination of the flow. Our geometry assumes, following
\citet{2026ApJ...996L..15O}, that the shear is purely along the LoS. If regions
B and C actually move at an angle to the LoS, the true tangential velocity is
larger still --- an effect distinct from, and additional to, the averaging
correction of Section~\ref{sec:sysU} --- and the interface is inclined, so that
a sky region of a given angular size samples a longer path across the layer.
The latter reduces the inferred $d$ and hence $\mu$ quadratically, while the
former raises $\mu$ linearly, so the net effect is to lower $\mu$. The
interaction times $T$ carry a further uncertainty of a factor of a few through
the assumed extents $R$ and $L$, but $\mu \propto T^{-1}$ only, so this is
subdominant.

The temperature adopted for region A also deserves comment. In the main
analysis of \citet{2026ApJ...996L..15O} it was frozen at $kT = 8.00$~keV, since
the region suffers from limited statistics and from spectral leakage from the
brighter region B. When left free, the fit prefers $kT = 5.75^{+1.05}_{-0.79}$
keV, which lowers $\msp$ to $\left( 2.0^{+1.0}_{-0.6} \right) \times 10^{3}$
\visc\ and raises $\mu_{\rm A}/\msp$ to $1.7^{+0.7}_{-1.0}$, still consistent
with unity. Because $\msp \propto T^{5/2}$, the comparison on the hot side is
sensitive to this choice at the factor-of-two level, and the error on $\msp$
then dominates over that on $\mu_{\rm A}$. Neither problem affects the cool
side, whose temperature is measured rather than assumed, and measured well
enough that $\msp$ contributes less to the error on $\mu_{\rm B}/\msp$ than
$\mu_{\rm B}$ itself does.

Taken together, these considerations place the result in the same qualitative
regime as the imaging constraints (Section~\ref{sec:intro}): the effective
viscosity of the ICM is at
most of order the Spitzer value, and quite possibly far below it. Even under
the maximal averaging correction of Section~\ref{sec:sysU}, which would raise
the estimates by a factor of two, the two sides of the front give
$\mu/\msp \approx 13$ and $1.5$; and that correction is itself disfavored by the
agreement between the mean velocities of regions B, C and D. The value of the
present exercise is not that it improves on the existing limits, but that it
reaches them by an entirely different route, using velocities rather than
densities.

\subsection{Implications for the Stability of the Front}

A KH perturbation of wavelength $\lambda$ at the interface grows at a rate
$\sim U/\lambda$ and is damped viscously at a rate $\sim \nu/\lambda^{2}$, so
damping wins below
\begin{equation}
\lambda_{\rm visc} \sim \frac{\nu}{U} \, .
\label{eq:lvisc}
\end{equation}
Taking the fiducial $U = 535$~\kms\ and the kinematic viscosities of
Table~\ref{tab:results} gives $\lambda_{\rm visc} \lesssim 5$~kpc on the cool
side and $\lesssim 13$~kpc on the hot side, these being upper limits because
the $\nu$ are. The estimate is insensitive to the systematic of
Section~\ref{sec:sysU}: since $\nu \propto U$, the ratio $\nu/U$ does not
change if $U$ is revised upward.

These figures say that our constraint leaves KH modes free to grow on the
scales at which structure is actually seen along the A3667 front, tens of kpc
and above \citep{2017MNRAS.467.3662I}. Viscosity alone therefore cannot be what
keeps the interface sharp, which points to the magnetic interpretation of the
front's sharpness \citep{2001ApJ...549L..47V,2011ApJ...743...16Z}, for which
\citet{2026ApJ...996L..15O} derive $B \gtrsim 5\ \mu$G from the same velocity
jump. The two mechanisms are not mutually exclusive: the required field strength
would be reduced if part of the stabilization were viscous, but our limit does
not permit viscosity to do the job on its own.

We stress that the quantity constrained here is an effective, isotropic
viscosity. In a magnetized ICM the momentum transport is anisotropic
\citep{1965RvPP....1..205B}, and the effective isotropic value inferred from a
macroscopic shear layer depends on the field geometry relative to the flow. At
a cold front the field is expected to be stretched parallel to the interface by
the shear itself \citep{2011ApJ...743...16Z}, a configuration in which the
Braginskii viscosity transports momentum poorly across the interface:
\citet{2015ApJ...798...90Z} find it to be far less effective than an isotropic
viscosity of the same nominal strength at suppressing KH instabilities on
sloshing cold fronts, for exactly this reason. A
kinematically measured $\mu$ that is comparable to or below $\msp$ is therefore
naturally accommodated.

\subsection{Prospects}
\label{sec:prospects}

The method is limited almost entirely by angular resolution, through
$\mu \propto d^{2}$. XRISM/Resolve, with a half-power diameter of
$\sim 1\farcm 3$ and $30''$ pixels, can only tell us that the shear layer of
A3667 is narrower than $\sim 100$~kpc. A future microcalorimeter with a
$\sim 5''$ mirror, such as the X-IFU on {\it Athena}
\citep{2023ExA....55..373B}, would resolve $d$ down to $\sim 5$~kpc at the
distance of A3667. If the layer remained unresolved at that scale, the same
analysis would yield $\mu \lesssim 15$~\visc, i.e., $\lesssim 0.02\, \msp$.
That is where the kinematic route would first match the tightest imaging
limits, including the one \citet{2017MNRAS.467.3662I} obtained for this very
front --- but reached directly from the velocity field, and with the quadratic
scaling meaning that every further factor in resolution buys two in $\mu$.
Applying the method to several cold fronts spanning a
range of temperatures and densities would then test the predicted scaling of
the effective viscosity, and thereby discriminate between Coulomb collisions
and microscale plasma instabilities as the agent that sets the momentum
transport in the ICM.

\section{Summary}
\label{sec:summary}

We have proposed a kinematic method for measuring the effective viscosity of
the ICM, in which the thickness of the viscous boundary layer that develops at
a sliding contact discontinuity is read off from spatially resolved
high-resolution spectroscopy. Treating the interface as Stokes' first problem
gives $\nu \approx d^{2}/(4T)$, where $d$ is the layer thickness and $T$ the
duration of the interaction.

Applying this to the XRISM/Resolve observation of the prototypical cold front
in A3667 \citep{2026ApJ...996L..15O}, where the velocity transition across the
front is complete within the $\approx 100$~kpc width of the region just inside
it --- the only region whose line is anomalously broad, with
$\sigma_{\rm B} = 424$~\kms\ --- we take $d \lesssim 100$~kpc for the shear layer
and obtain dynamic viscosities of
$\mu \lesssim \left( 5.2^{+1.6}_{-1.5} \right) \times 10^{3}$~\visc\ inside the
front and $\lesssim \left( 3.3^{+1.0}_{-1.0} \right) \times 10^{3}$~\visc\
outside it, where the errors propagate the measured uncertainty on the velocity
jump. These correspond to $6.4^{+2.2}_{-2.2}$ and $0.72^{+0.23}_{-0.21}$ times
the respective Spitzer
values, so that the ICM viscosity is at most a few times Spitzer and may be far
smaller.

The shear velocity entering these estimates deserves care, since the quantity
that the model requires is the far-field relative velocity while the quantity
that is measured is a difference of region-averaged velocities, which is
smaller. Correcting for this raises $\mu$ linearly, by at most a factor of two
for our geometry, but the correction is disfavored by the close agreement
between the mean velocities of the region just inside the front and of the cool
gas beyond it, which indicates that most of the gas in that region has not in
fact been dragged.

The constraint is currently set by the angular resolution of XRISM rather than
by the quality of its spectra. Because $\mu \propto d^{2}$, an instrument
combining microcalorimeter spectroscopy with arcsecond imaging would improve it
by orders of magnitude, and would turn the viscosity of the ICM from an
inference about images into a direct measurement of a velocity field.

\begin{acknowledgments}
We thank the XRISM collaboration for making these measurements possible.
We acknowledge support from JSPS KAKENHI Grant Nos. 23H04899 and 25H00672
(Y.F.).
\end{acknowledgments}

\bibliographystyle{aasjournalv7}
\bibliography{viscosity_a3667}

\end{document}